\documentclass[nofootinbib, letterpaper, floatfix, preprintnumbers, twocolumn]{revtex4}  

\pdfoutput=1

\usepackage{amsmath,amssymb}
\usepackage{graphicx} 
\usepackage{epstopdf} 
\usepackage{slashed}
\usepackage{color}
\usepackage{multirow}
\usepackage{footmisc}
\usepackage{hyperref}
\usepackage{subfigure}
\usepackage{snapshot}
\usepackage{blindtext}
\usepackage{hepunits}

\usepackage{array}
\newcolumntype{L}[1]{>{\raggedright\let\newline\\\arraybackslash\hspace{0pt}}m{#1}}
\newcolumntype{C}[1]{>{\centering\let\newline\\\arraybackslash\hspace{0pt}}m{#1}}
\newcolumntype{R}[1]{>{\raggedleft\let\newline\\\arraybackslash\hspace{0pt}}m{#1}}

\DefineFNsymbols*{lamport}{\dagger\ddagger\S\P\|{**}{\dagger\dagger}{\ddagger\ddagger}}

\newcommand{\be}{\begin{eqnarray*}}
\newcommand{\ee}{\end{eqnarray*}}

\newcommand{\bee}{\begin{eqnarray}}
\newcommand{\eee}{\end{eqnarray}}
\newcommand{\beeq}{\begin{equation}}
\newcommand{\eeeq}{\end{equation}}

\newcommand{\fb}{{\text{fb}}}

\begin{document}

\title{A 750 GeV graviton from holographic composite dark sectors}
%
%

\begin{abstract}
	We show that the $750\,$ GeV di-photon excess can be interpreted as a spin-2 resonance arising from a strongly interacting dark sector featuring some departure from conformality. This spin-2 resonance has negligible couplings to the SM particles, with the exception of the SM gauge bosons which mediate between the two sectors. We have explicitly studied the collider constraints as well as some theoretical bounds in a holographic five dimensional model with a warp factor that deviates from AdS$_5$. In particular, we have shown that it is not possible to decouple the vector resonances arising from the strong sector while explaining the di-photon anomaly and keeping the five dimensional gravity theory under perturbative control. However, vector resonances with masses around the $\TeV$ scale can be present while all experimental constraints are met. 

\end{abstract}

\author{Adri\'an Carmona}
\email{adrian.carmona@cern.ch}
\affiliation{CERN, Theoretical Physics Department,\\ 1211 Geneva 23, Switzerland}

\preprint{CERN-TH-2016-073}

\maketitle

\section{Introduction}
\label{sec:intro}

The discovery of the Higgs boson by the ATLAS and CMS collaborations at the Large Hadron
Collider (LHC) marked the beginning of a new era in high energy  physics. Indeed, the finding of the long-sought particle offers us the unique opportunity to start testing the origin of electroweak symmetry breaking (EWSB). This means that we could be closer than ever to understand some extremely important unsolved puzzles in particle physics, like the large hierarchy between the electroweak and the Planck scales,  the origin of fermion masses or even what lies behind Dark Matter (DM). The situation has become even more thrilling after the announcement by both ATLAS and CMS collaborations of a tantalizing hint of a new resonance in di-photon production at masses around $\sim 750\,$GeV \cite{ATLAS-CONF-2015-081, CMS-PAS-EXO-15-004, CMS-PAS-EXO-16-018}. Since the exciting news awakened the feverish imagination of theorists, we have witnessed a plethora of papers exploring possible explanations of the reported anomaly. However, for several reasons, the spin-2 possibility  has been largely unexplored (see e.g \cite{Han:2015cty, Giddings:2016sfr, Sanz:2016auj, Falkowski:2016glr, Liu:2016mpd, Hewett:2016omf}). One of the reasons for this oblivion is that traditional \emph{vanilla} explanations in terms of Kaluza-Klein (KK) gravitons face several problems for such light masses, since they favor either universal couplings to the Standad Model (SM) content or very small $\gamma \gamma$ branching ratios, which are not viable phenomenologicaly. In addition, unless large localized curvature terms make the spin-2 resonance much lighter than the rest of the KK spectrum, the constraints resulting from electroweak precision tests (EWPT) clearly exclude such scenarios. Moreover, it is known that the presence of such terms can easily turn the radion into a ghost\,\cite{Davoudiasl:2003zt, Luty:2003vm}, questioning the viability of these setups. In this letter we will explore an interesting possibility where the reported $750\,\GeV$ resonance may arise from a holographic strongly interacting dark sector. We will show that in  models where the strong sector features some deformation of conformality, parametrized in the five dimensional (5D) framework by a modified background, a light graviton can naturally explain the observed anomaly while still fulfilling all other experimental constraints arising from collider searches or EWPT. Moreover, we will demonstrate that all this can be done without introducing a too large gap between the masses of the KK graviton and the rest of the KK spectrum, which will allow to have perturbativity under control in the 5D gravity theory and avoid the emergence of a radion ghost. In addition, we will show that in these models there is a beautiful interplay between the dark sector (possibly explaining part of the observed relic abundance) and the collider phenomenology of the KK vectors. Therefore, measuring the properties of the hypothetical particle
, in case its existence is confirmed,   will definitively help to answer if it is related  to the origin of EWSB or rather with other fundamental puzzles in particle physics, like the  origin of DM. 

The article is organised as follows: in Section~\ref{sec:setup} we introduce the original theoretical motivation and the concrete 5D framework where all computations will be performed. This will also serve us to introduce notation and the input parameters of the theory. In Section~\ref{sec:pheno} we will examine in detail the phenomenological consequences of the proposed setups, studying in detail the interplay between EWPT, the different collider searches and role played by the DM candidates.  Finally, we conclude in Section~\ref{sec:conc}.

\section{Theoretical Motivation and Setup}
\label{sec:setup}
Trying to address the hierarchy problem has provided us a better understanding of the SM as well as stimulating theoretical constructions  like supersymmetry, composite Higgs models, technicolor or models with warped extra dimensions. However, the multiplication of negative results for such theories  has propelled alternative ways of thinking about new physics, disconnecting it e.g. from the electroweak scale. One particular example is the case of DM, where some of these theoretical constructions have been used with the goal of explaining its origin with no   deep connection with the  electroweak scale, see e.g. \cite{Kilic:2009mi, Kilic:2010et, Bai:2010qg, Antipin:2015xia, Carmona:2015haa, Redi:2016kip}.\,\footnote{All these models explore the possibility of having a strongly coupled sector not involved in EWSB. While in \cite{Kilic:2009mi, Kilic:2010et, Bai:2010qg} and \cite{Carmona:2015haa}  the strong sector only talks to the SM via gauge interactions, in \cite{Antipin:2015xia, Redi:2016kip} Yukawa interactions are sometimes  allowed. Moreover, Ref.\,\cite{Carmona:2015haa} focus on the effective holographic description of such scenarios.} In the case of models with warped extra dimensions or theirs strongly coupled duals, this  more modest and pragmatic approach has some advantages, for typical problems associated with these scenarios are turned into advantages once the hierarchy problem is left unsolved. For instance, in Ref.\,\cite{Carmona:2015haa}  the most minimal examples where the full SM (including the Higgs boson) is extended with a strongly-interacting composite sector  delivering  pseudo Nambu-Goldstone bosons (pNGBs) as natural DM candidates were studied. In this letter we are going to explore the possibility that the first spin-2 resonance arising in their holographic constructions can explain the $750\, \GeV$ di-photon anomaly.\,\footnote{For other examples of spin-2 resonances arising from strongly interacting dark sectors, see e.g. \cite{Lee:2013bua, Lee:2014caa}.} There has been some recent studies on the possibility of interpreting the putative $750\,\GeV$ resonance as a KK graviton arising from extra dimensional setups \cite{Han:2015cty, Giddings:2016sfr, Sanz:2016auj, Falkowski:2016glr, Liu:2016mpd, Hewett:2016omf} but only  Refs.\,\cite{Han:2015cty} and \cite{Falkowski:2016glr} considered the case where the whole SM matter content is UV localized and only gauge bosons are allowed to propagate into the bulk. However, none of them considered the effect of the vector resonances, which were ignored or lifted to $\sim 3-4\,\TeV$  without considering the implications on the consistency of the 5D theory or the radion dynamics. Moreover, we will study the more general case where deformations of conformality in the strong sector are allowed, which is parametrized in the 5D theory by a more general warp factor. This will increase the generality of the approach and will improve the agreement with EWPT and collider constraints.

\begin{figure}[!h]
	\begin{center}
		\includegraphics[width=\columnwidth]{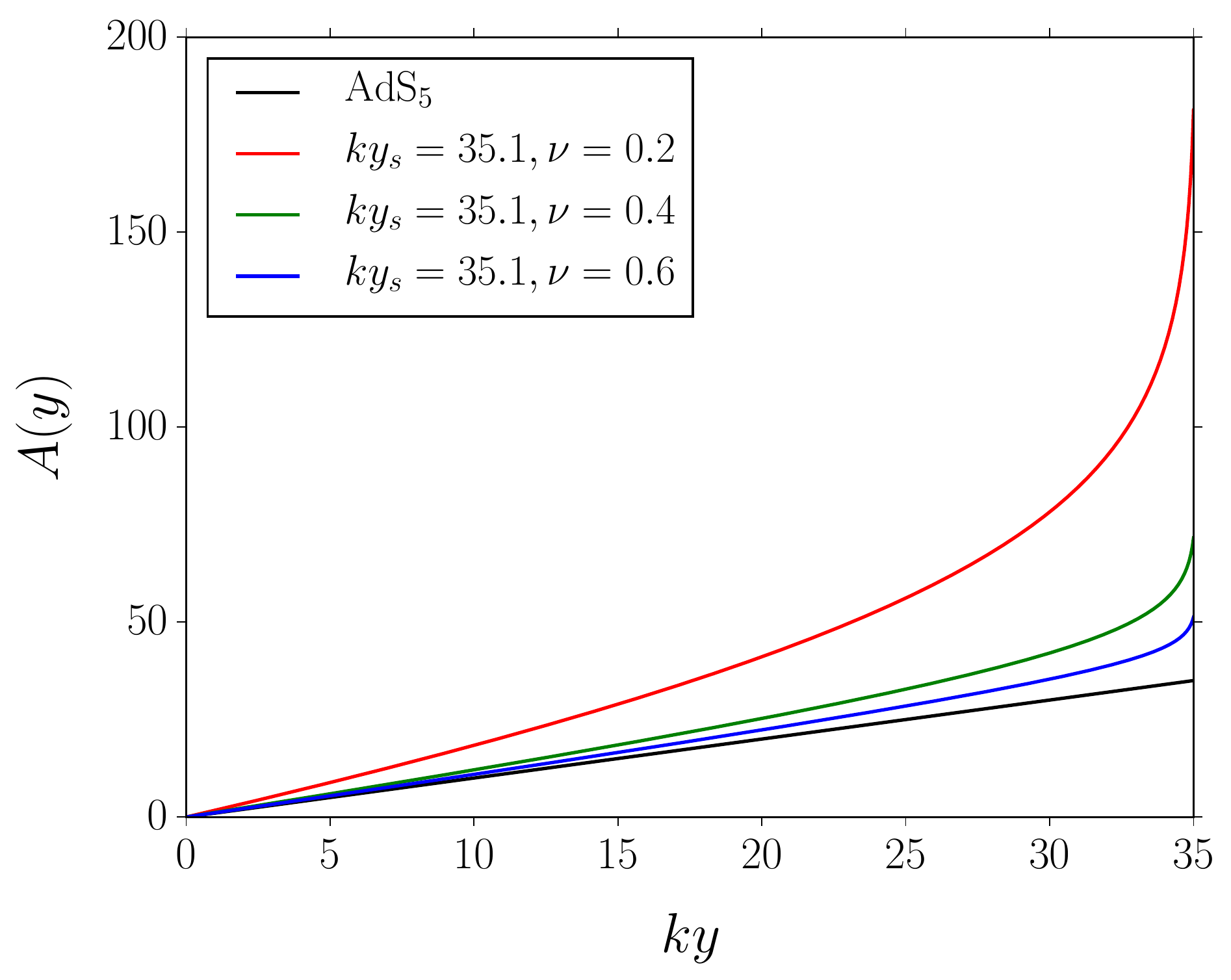}
		\caption{Warp factor as defined in eq.~(\ref{eq:metric}) for $ky_1=35$, $ky_s=35.1$ and different values of $\nu$. We also show the AdS$_5$ case for comparison, which corresponds to the limits $\nu\to \infty$ or $k y_s\to \infty$.}
		\label{fig:metric}
	\end{center}
\end{figure}

We consider a slice of extra dimension with the following metric 
\begin{equation}
	ds^2=e^{-2A(y)}\eta_{\mu\nu}dx^{\mu}dx^{\nu}-dy^2,
\end{equation}
where the warp factor is given by \cite{Cabrer:2009we, Cabrer:2010si, Cabrer:2011fb, Cabrer:2011vu, Carmona:2011ib, Cabrer:2011qb}
\begin{eqnarray}
	A(y)=ky-\frac{1}{\nu^2}\mathrm{log}\left(1-\frac{y}{y_s}\right),
	\label{eq:metric}
\end{eqnarray}
and the extra dimension is parametrized by the coordinate $y\in[0,y_1]$,   bounded by two fixed points or branes, corresponding to $y=0$ (UV brane) and $y=y_1$ (IR brane). On the other hand, $y_s>y_1$ represents the position of the singularity responsible for the  deformation of conformality, with the AdS$_5$ case being recovered  in the limits $y_s\to \infty$ or $\nu\to \infty$. We show in Figure\,\ref{fig:metric} the warp factor for different values of $\nu$ for $ky_1=35$ and $ky_s=35.1$, as well as for the AdS$_5$ case. We  trade $y_s$ by the curvature radius at the IR brane, given (in units of $k$) by 
\begin{eqnarray}
	k L_1=\frac{\nu^2 k (y_s-y_1)}{\sqrt{1-2\nu^2/5+2\nu^2	k(y_s-y_1)+\nu^4k^2(y_s-y_1)^2}},
\end{eqnarray}
where $0.1\lesssim k L_1 \lesssim 1$. The value of $y_1$ can be fixed by chosing different values of the UV/IR hierarchy $A(y_1)$. The AdS$_5$ limit corresponds to $A(y_1)\sim 36$ and $kL_1\to1$. 

In the transverse-traceless gauge, the spin-2 gravitational excitations are parametrized by the tensor fluctuations of the metric $\eta_{\mu\nu}\to \eta_{\mu\nu}+\kappa_5 h_{\mu\nu}(x,y)$, where $\partial^{\mu}h_{\mu\nu}=h_{\alpha}^{\alpha}=0$ and $\kappa_5=2 M_5^{-3/2}$, with $M_5$ the 5D Planck mass. The graviton KK expansion reads
\begin{eqnarray}
	h_{\mu\nu}(x,y)=\sum_n h^{(n)}_{\mu\nu}(x)f^{(n)}_{h}(y),
\end{eqnarray}
where $f^{(n)}_h$ satisfy
\begin{eqnarray}
	(e^{-4A(y)}f^{(n)\prime}_h(y))^{\prime}+e^{-2A(y)}m^{(n)2}_h f_{h}^{(n)}(y)=0,
\end{eqnarray}
and 
\begin{eqnarray}
	0&=&f_{h}^{(n)\prime}(0)+\kappa_0 k^{-1} m^{(n)2}_h f_h^{(n)}(0)\\
	 &=&e^{-2 A(y_1)}f_h^{(n)\prime}(y_1)-\kappa_1 k^{-1} m_h^{(n)2} f_h^{(n)}(y_1),\nonumber
\end{eqnarray}
in presence of possible localized curvature terms \cite{Davoudiasl:2003zt}. These profiles are normalized  
\begin{eqnarray}
	\int_0^{y_1}dy\, e^{-2A(y)} f_{h}^{(n) 2}[1+\delta(y)\frac{\kappa_0}{k} + \delta(y-y_1)\frac{\kappa_1}{k}]=1,
\end{eqnarray}
in such a way that
\begin{eqnarray}
	\bar{M}_{\rm Pl}^2=M_5^3\int_0^{y_1}d y e^{-2A(y)}[1+\delta(y)\frac{\kappa_0}{k}+\delta(y-y_1)\frac{\kappa_1}{k}],
	\label{eq:m5match}
\end{eqnarray}
where $\bar{M}_{\rm Pl}=2.4\times  10^{18}\,\GeV$ is the four-dimensional reduced Planck mass. 

In the spirit of the models considered in Ref.\,\cite{Carmona:2015haa}, we assume that only the SM gauge bosons propagate into the bulk of the extra dimension, with the full SM matter content being localized at the UV brane.\,\footnote{Considering some relatively high new physics scale at the UV brane alleviating the hierarchy problem, would not change the picture, provided the light degrees of freedom remain those of the SM (assuming therefore some moderate fine-tuning). }  In addition, we also assume that the bulk of the extra dimension respects a larger gauge group, like e.g. $SU(3)\times SU(3)\times U(1)_X$ or $SU(3)\times SU(2)_1\times SU(2)_2\times U(1)_X$, which delivers some dark pNGBs $A_{5}^{\hat{a}}(x)$.  We expect therefore additional  spin-1 KK resonances in addition to the usual electroweak vector ones. However, since they do not couple to the SM, they will play no role in the current phenomenological analysis.  The additional scalars, on the other hand,  will have sizable couplings to the electroweak vector resonances, for they are all localized towards the IR brane, making the latter to decay almost exclusively to these scalars, as was explicitly shown in Ref.\,\cite{Carmona:2015haa}. At any rate,  the only relevant input from such constructions in the current study is the introduction of a large invisible width for the electroweak vector resonances, that makes the bounds from color octet searches the leading ones.  


The KK expansion of the SM gauge bosons reads $\mathbb{A}_{\mu}(x,y)=\sum_nf_{\mathbb{A}}^{(n)}(y)\mathbb{A}_\mu^{(n)}(x,y)$ where $\mathbb{A}_{\mu}=A_{\mu},Z_{\mu},W^{\pm}_{\mu},G_{\mu}^a$. Their profiles satisfy the following bulk equations of motion 
\begin{eqnarray}
	(e^{-2A(y)}f^{(n)\prime}_{\mathbb{A}})^{\prime}+m^{(n)2}_{\mathbb{A}} f_{\mathbb{A}}^{(n)}=0,
\end{eqnarray}
and boundary conditions
\begin{eqnarray}
	&&f_{\mathbb{A}}^{(n)\prime}(y_1)=0=f_{A}^{(n)\prime}(0)=f_{G}^{(n)\prime}(0)\\
	&=&\left.[\partial_y-\frac{v^2}{4}(g_5^2+g_5^{\prime 2})]f_{Z}^{(n)}\right|_{y=0}=\left.[\partial_y-\frac{v^2}{4}g_5^2]f_{W}^{(n)}\right|_{y=0}.\nonumber
\end{eqnarray}
In order to be slightly more general, we also allow for localized UV gauge kinetic terms (KT), $\kappa_S^2 y_1$ and $\kappa_{EW}^2 y_1$, that  change the UV boundary conditions above by $\partial_y\to \partial_y +m^2_{\mathbb{A}} \kappa_{S,EW}^2 y_1$. These KT also change the normalization conditions for the different gauge profiles
\begin{eqnarray}
	\int_0^{y_1}dy f_{\mathbb{A}}^{(n)2}(y)+f_{\mathbb{A}}^{(n)2}(0)\kappa_{S,EW}^2 y_1=1.
\end{eqnarray}
However, in practice, these terms just basically change the matching of the 5D gauge couplings
\begin{eqnarray}
	g_5=g \sqrt{y_1(1+\kappa_{EW}^2)},\quad g_{5s}=g_s\sqrt{y_1(1+\kappa_{S}^2)},
\end{eqnarray}
whereas the ratio
\begin{eqnarray}
	g_{5}^{\prime}/g_5\sim g^{\prime}/g=\tan\theta_W
\end{eqnarray}
remains unchanged, for we have chosen identical KT for $SU(2)_L$ and $U(1)_Y$. Besides the gauge and gravitational kinetic terms $\kappa_{S,EW}^2$ and $\kappa_{0,1}$, we have five additional input parameters in the theory $M_5, A(y_1), \nu, k$ and $kL_1$. We can fix $M_5$  using  $\bar{M}_{Pl}$ and equation (\ref{eq:m5match}), whereas $m_{h}^{(1)}=750\, \GeV$ allow us to remove e.g. $\kappa_1$. For simplicity, we will chose $\kappa_0=0=\kappa_{\rm EW}$   leaving us in total with only  four parameter  $\{\nu, k L_1, m_{\rm KK}, \tilde{\kappa}, A(y_1) \}$, where we have traded $k$ for the first vector KK mass $m_{\rm KK}$, and defined $\tilde{\kappa}=\sqrt{1+\kappa_{S}^2}$.

The KK-graviton interactions are given by
\begin{eqnarray}
	&&\mathcal{L}\supset -\frac{\kappa_5}{2}\sum_{n=1}^{\infty}\sqrt{g_{\rm UV}}\Theta_{\mu\nu}^{\rm UV}(x)f_h^{(n)}(0)h_{\rho\gamma}^{(n)}(x)\eta^{\mu\rho}\eta^{\nu\gamma}\\
	&& -\frac{\kappa_5}{2}\sum_{n=1}^{\infty}\int_0^{y_1}dy\sqrt{g}e^{2A(y)} \Theta_{\mu\nu}(x,y)f_h^{(n)}(y)h_{\rho\gamma}^{(n)}(x)\eta^{\mu\rho}\eta^{\nu\gamma}\nonumber 
\end{eqnarray}
where  $\sqrt{g}=e^{-4 A(y)}$ and $\sqrt{g_{\rm UV}}=1$ are the square root of the determinant of the 5D and the UV-localized metrics, respectively, whereas
\begin{eqnarray}
	\Theta_{\mu\nu}&=&-\frac{2}{\sqrt{g}}\frac{\delta (\sqrt{g}\mathcal{L}_{\rm matter})}{\delta g^{\mu\nu}}=-2\frac{\delta \mathcal{L}_{\rm matter}}{\delta g^{\mu\nu}}\nonumber\\
&&+g_{\mu\nu}\mathcal{L}_{\rm matter},\end{eqnarray}
and
\begin{eqnarray}
	\Theta_{\mu\nu}^{\rm UV}=-2\frac{\delta \mathcal{L}_{\rm matter}^{\rm UV}}{\delta \eta^{\mu\nu}}+\eta_{\mu\nu}\mathcal{L}_{\rm matter}
\end{eqnarray}
are the bulk and UV-localized stress-energy tensors. We can neglect the last piece in the stress-energy tensors above, for the graviton is in our \emph{gauge} traceless, considering only 
\begin{eqnarray}
	T_{\mu\nu}=-2\frac{\delta \mathcal{L}_{\rm matter}}{\delta g^{\mu\nu}}, \quad T_{\mu\nu}^{\rm UV}=-2\frac{\delta \mathcal{L}_{\rm matter}^{\rm UV}}{\delta \eta^{\mu\nu}}.
\end{eqnarray}
We obtain therefore
\begin{eqnarray}
	T_{\mu\nu}^{\mathcal{A}}=e^{2 A(y)}F_{\mu\beta}^{\mathcal{A}}F_{\nu\gamma}^{\mathcal{A}}\eta^{\beta\gamma}.
	\label{eq:teng}
\end{eqnarray}
for $SU(3)_c\times SU(2)_L\times U(1)_Y$ gauge bosons, where $\mathcal{A}_{\mu}=G^a_{\mu},W^I_{\mu},B_{\mu}$.  Regarding the UV-localized SM sector, we obtain
\begin{eqnarray}
	T^{G\,\textrm{UV}}_{\mu\nu}&=&(\tilde{\kappa}^2-1) y_1 F_{\mu\beta}^{G}F_{\nu\gamma}^{G}\eta^{\beta\gamma},\\
	T^{H\,\textrm{UV}}_{\mu\nu}&=&-2D_{\mu}H D_{\nu}H,\\
	T_{\mu\nu}^{\Psi\,\textrm{UV}}&=&-i\bar{\Psi}D_{[\mu}\gamma_{\nu]}\Psi,
\end{eqnarray}
for fermions $\Psi$ and the Higgs doublet $H$, where $D_{\mu}$ is the usual SM covariant derivative and we have defined $D_{[\mu}\gamma_{\nu]}=D_{\mu}\gamma_{\nu}-D_{\nu}\gamma_{\mu}$. Since the KK graviton is exponentially peaked towards the IR brane, the interactions resulting from the above UV-localized terms are negligible  compared to the ones coming from (\ref{eq:teng}), so we will safely neglect them henceforth. 
\section{Phenomenological study}
\label{sec:pheno}
\begin{figure}[!h]
	\centering
	\includegraphics[width=\columnwidth]{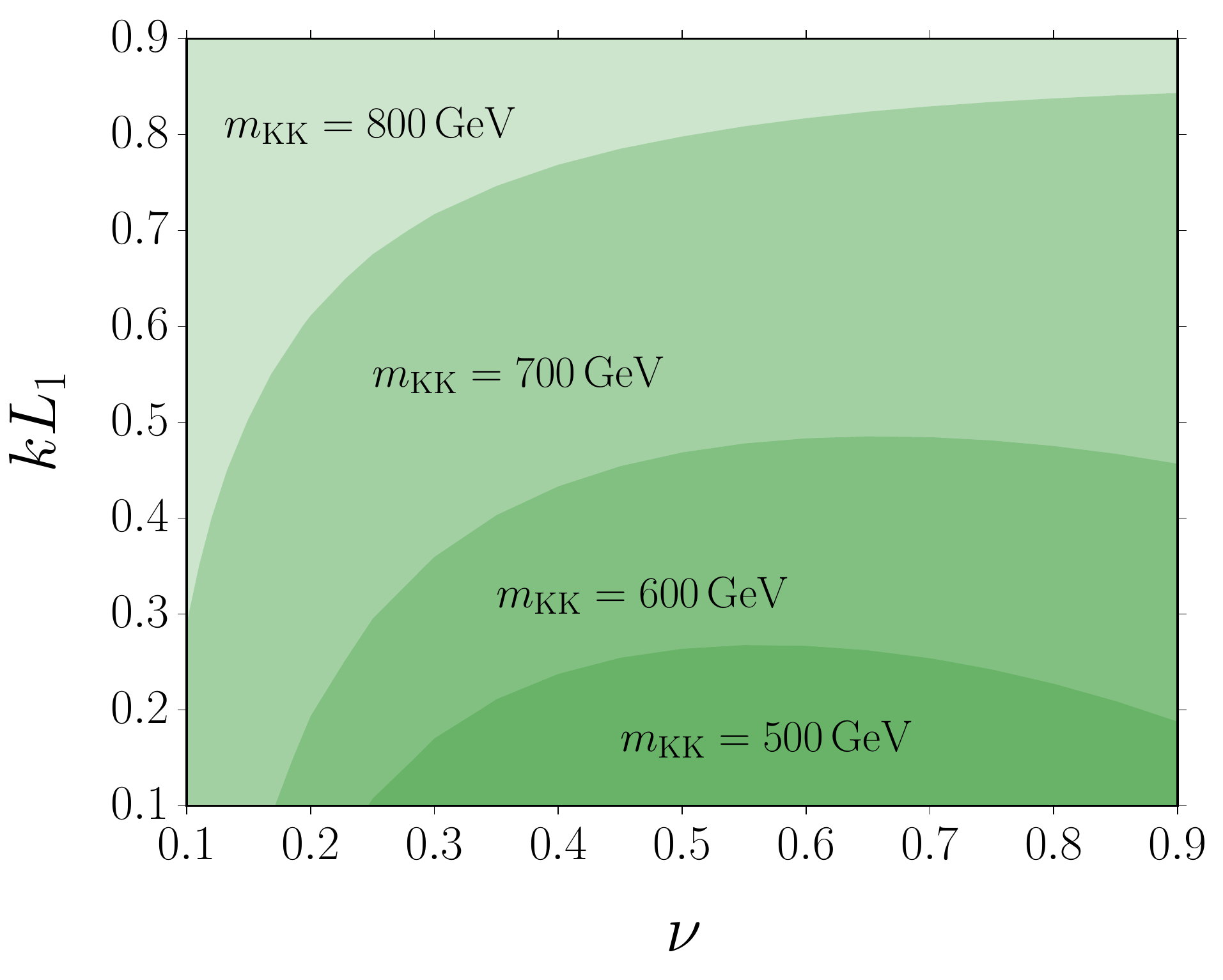}
	\caption{Constraints from EWPT at 95\% C.L. in the $\nu-kL_1$ plane for different values  of $m_{\rm KK}$ assuming $A(y_1)=37.5$. For each value of $m_{\rm KK}$, values of $\nu$ and $kL_1$ within the corresponding green region are allowed.}
		\label{fig:wyfit}
\end{figure}

\begin{figure}[!h]
	\centering
	\includegraphics[width=\columnwidth]{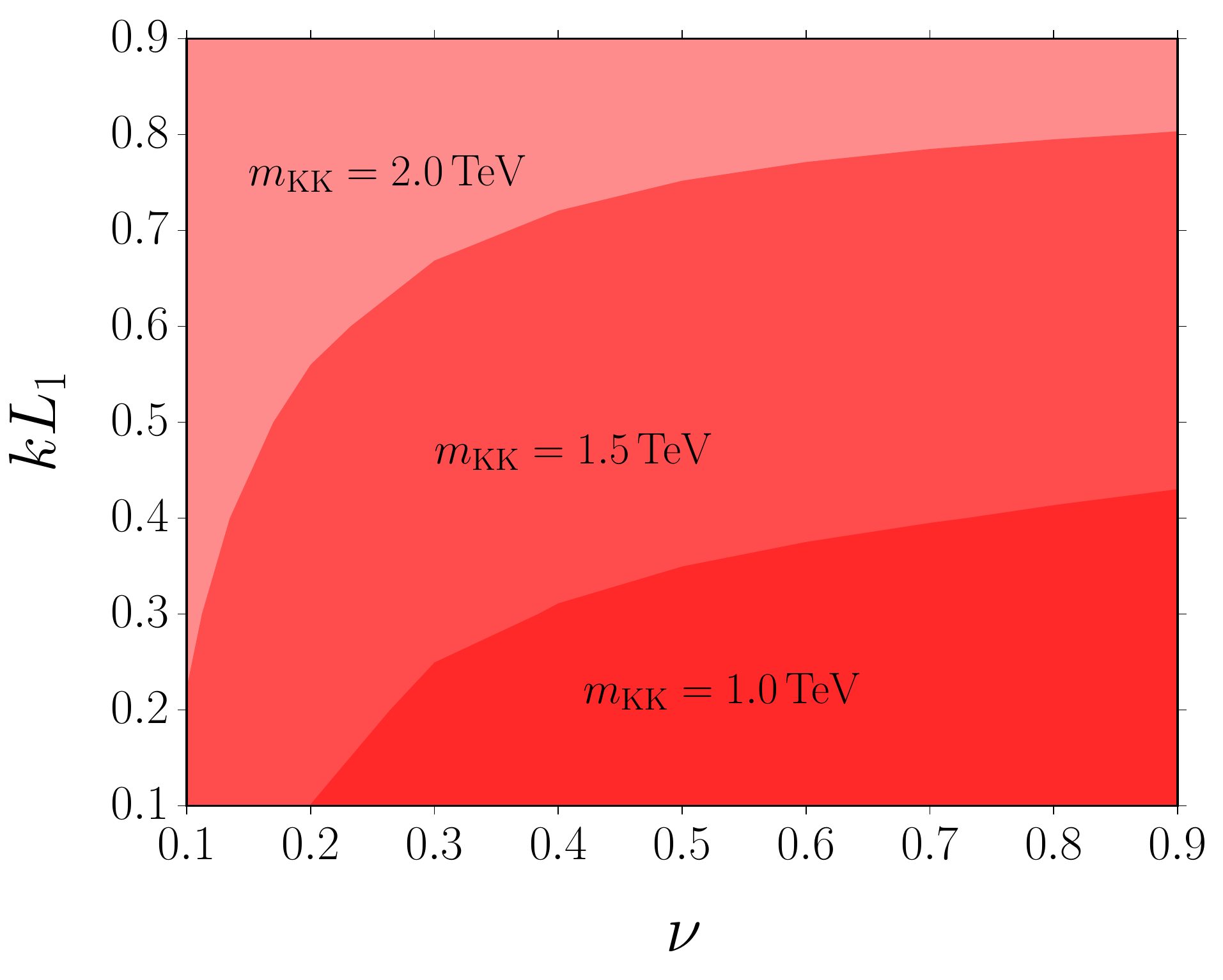}
	\caption{Excluded regions  for losing 5D perturbativity control in the $\nu-kL_1$ plane for different values  of $m_{\rm KK}$, assuming  $A(y_1)=37.5$ and $M_5L_1\lesssim 0.4$. For each value of $m_{\rm KK}$, values of $\nu$ and $kL_1$ within the corresponding red region are excluded. The RS limit $k L_1\to 1$ is also excluded for $m_{\rm KK}=2\,\TeV$.}
		\label{fig:M5L1}
\end{figure}

One of the first logical concerns of having a $750\,\GeV$ KK-graviton (which is not anomalously light compared with the rest of the KK spectrum), is the possible conflict with EWPT.  However, since the full SM matter content is localized on the UV brane and the extra dimension plays no role on EWSB,  the oblique paremeters $S$ and $T$ are zero at tree level, which alleviates enormously the pressure from EWPT. Therefore, the only relevant constraint in this regard are the volume suppressed  $W$ and $Y$ operators \cite{Barbieri:2004qk},
\begin{eqnarray}
	W=\frac{g^2 M_W^2}{2}\Pi^{\prime\prime}_{W_3 W_3}(0),\quad Y=\frac{g^{\prime 2} M_W^2}{2}\Pi_{BB}^{\prime\prime}(0),
\end{eqnarray}
which are given by \cite{Cabrer:2011fb}
\begin{eqnarray}
	Y=W=\frac{c_W^2 m_Z^2 }{y_1 }\int_0^{y_1}e^{2 A(y)}\left(y_1-y\right)^2.
\end{eqnarray}
We have performed an up-to-date fit to $W=Y$,\,\footnote{We thank Jorge de Blas for providing us the $\chi^2$ for the EW fit, which includes all the
observables considered in the analysis of \cite{delAguila:2011zs, deBlas:2013gla}, updated with the current experimental values.} and the allowed values at $95\%$\,C.L. are shown in Figure\,\ref{fig:wyfit} for different values of $m_{\rm KK}$ in the $\nu-kL_1$ plane, assuming the benchmark value $A(y_1)=37.5$, since it will provide the sought cross section for the di-photon anomaly. One could wonder of such choice since the hierarchy problem is not longer addressed by the extra dimension. However, we still want to have a 5D theory of gravity with a $\sim \TeV$ KK graviton and, as we will see $\TeV$ vector resonances,  so it is not surprising that we end up considering similar values to the original RS model.  We can readily see from the plot that large deformations of conformality are strongly preferred by the data, for only small values of $k L_1$ are allowed for low values of $m_{\rm KK}$. Still, once $m_{\rm KK}$ approaches $1\,\TeV$ the bulk of the parameter space leads to agreement with EWPT. It is then tempting to arbitrarily increase  the masses of the vector resonances in order to avoid their experimental constraints. However, since the KK graviton mass is fixed at $750\,\GeV$, this is only possible at the price of reducing the perturbativity in the 5D gravity theory. Indeed,  as can be seen from Figure\,\ref{fig:M5L1}, where we show the regions of the parameter space with $M_5 L_1\lesssim 0.4$ (since for arbitrary small values of this dimensionless parameter perturbative control in the 5D gravity theory is lost) for different values of $m_{\rm KK}$, masses around $2\,\TeV$ are already excluded for $A(y_1)=37.5$.\,\footnote{Note that allowing for $\kappa_0>0$ would increase this tension, since it would result in smaller values of $M_5$ with no effect on $L_1$ (see eq.\,\ref{eq:m5match}). Negative values of $\kappa_0$ (which would need to be bigger than some lower bound to avoid a negative kinetic term for the massless graviton --  in the RS case, $\kappa_0>-1$),  could in principle increase $M_5$. However,  this effect could only be significant at the price of a considerable fine-tuning arising from the cancellation of the bulk contribution ($1+\kappa_0\sim 0$ in RS).} These bounds can be relaxed by reducing the volume factor $A(y_1)$, but this can not be done indefinitely since this also reduces the KK graviton cross section, as we will see below. 

\begin{figure}[!h]
	\centering
	\includegraphics[width=\columnwidth]{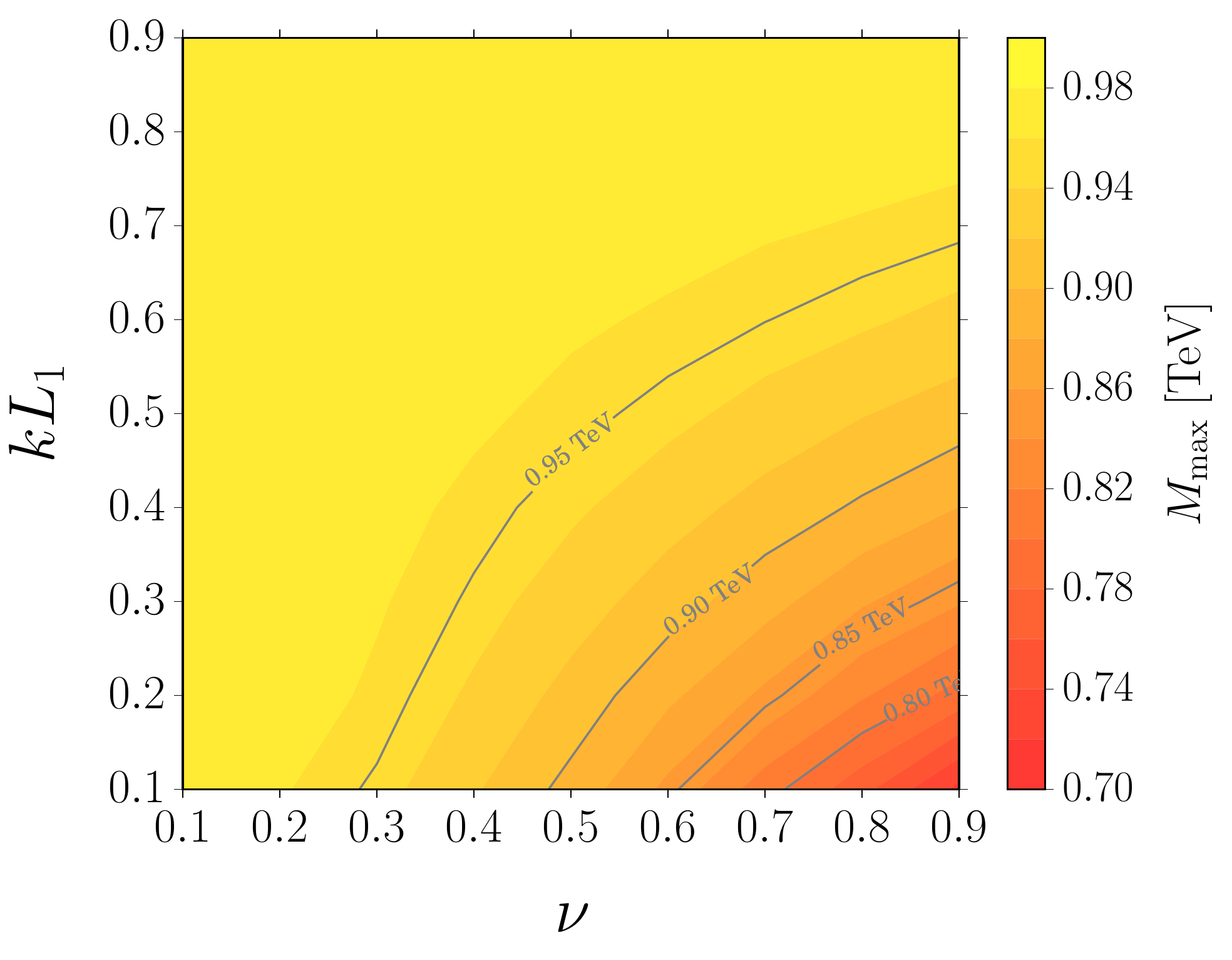}
	\caption{Maximum value of $m_{\rm KK}$, $M_{\rm max}$,  as a function of $\nu$ and $kL_1$ for $A(y_1)=37.5$. This value has been obtained requiring the absence of a radion ghost.}
		\label{fig:MaxM}
\end{figure}

Moreover, the size of the required graviton KT to produce a spectrum where $m_{\rm KK} \gtrsim m_{h^{(1)}}$ remains another source of concern. In the absence of such terms a KK graviton of $750\,\GeV$ would require KK gauge resonances with  masses $\sim 500\,\GeV$, since the ratio $m_{h^{(1)}}/m_{g^{(1)}}$ is fixed to $\approx 1.6$ (both in the RS case and for sizable deformations of conformality). Indeed, as it was already pointed out in \cite{Davoudiasl:2003zt, Luty:2003vm}, the presence of such terms produce a negative contribution to the radion KT, that can at some point turn it into  a ghost.  If we perform a similar analysis to the one carried out in \cite{Davoudiasl:2003zt} for the model at hand, we obtain that this will happen when
\begin{eqnarray}
Z_r=1-3 \kappa_1/k e^{-2A(y_1)} F^2(y_1) X_F^{-1}<0
	\label{eq:zr}
\end{eqnarray}
where $F(y)$ is the radion profile\,\footnote{We assume $\partial_y(e^{-2A(y)}F(y))=0$ boundary conditions on both branes.} (for more details see e.g. \cite{Cabrer:2010si, Cabrer:2011fb, Megias:2015ory}) and we have defined
\begin{eqnarray}
	X_F&\equiv&\int_0^{y_1}dy e^{-2A(y)}F^2(y)\left[6\phantom{\frac{1}{2}}\right.\nonumber\\
	&&+\left.\frac{36^2}{2\beta^2 W^2}\left(\frac{F^{\prime}}{F}-2A^{\prime}\right)^2\right],
\end{eqnarray}
with
\begin{eqnarray}
	\beta(\phi)W(\phi)&=&-6\sqrt{6}e^{\nu\phi/\sqrt{6}}k\nu,\\
	\phi(y)&=&-\sqrt{6}/\nu\log[\nu^2k(y_s-y)].
\end{eqnarray}
In the RS case, we get 
\begin{eqnarray}
	Z_r=1-\frac{1}{2}\kappa_1e^{2k y_1}\left[k\int_0^{y_1}dy\, e^{2ky}\right]^{-1}=1-\kappa_1,
\end{eqnarray}
which leads to $\kappa_1\le 1$, after imposing the absence of a radion ghost.  In general, this  bound on $\kappa_1$ can be translated into an upper bound on $m_{\rm KK}$ for any value of $k L_1$, $\nu$ and $A(y_1)$, by using  equation (\ref{eq:zr}). In Figure\,\ref{fig:MaxM}, we show a contour plot for this value, $M_{\rm max}$, in the $\nu-kL_1$ plane for $A(y_1)=37.5$. In the RS case, we obtain $M_{\rm max}\approx 1\,\TeV$. Finding if such bound can be somehow alleviated or it is an unavoidable constraint is an interesting theoretical puzzle \emph{per se}. However, if the appearance of a $750\, \GeV$ resonance is eventually confirmed, it will become a much more relevant question. Since we are not aware of any solution to this issue at the moment, we consider the limits from Figure\,\ref{fig:MaxM} $m_{\rm KK}\lesssim 1\,\TeV$ to be definitive. Therefore, we will consider $m_{\rm KK}=0.9\,\TeV,~0.95\,\TeV$ and $1\,\TeV$, even though the whole parameter space of the latter will be marginally excluded.

\begin{figure}[!h]
	\centering
	\includegraphics[width=\columnwidth]{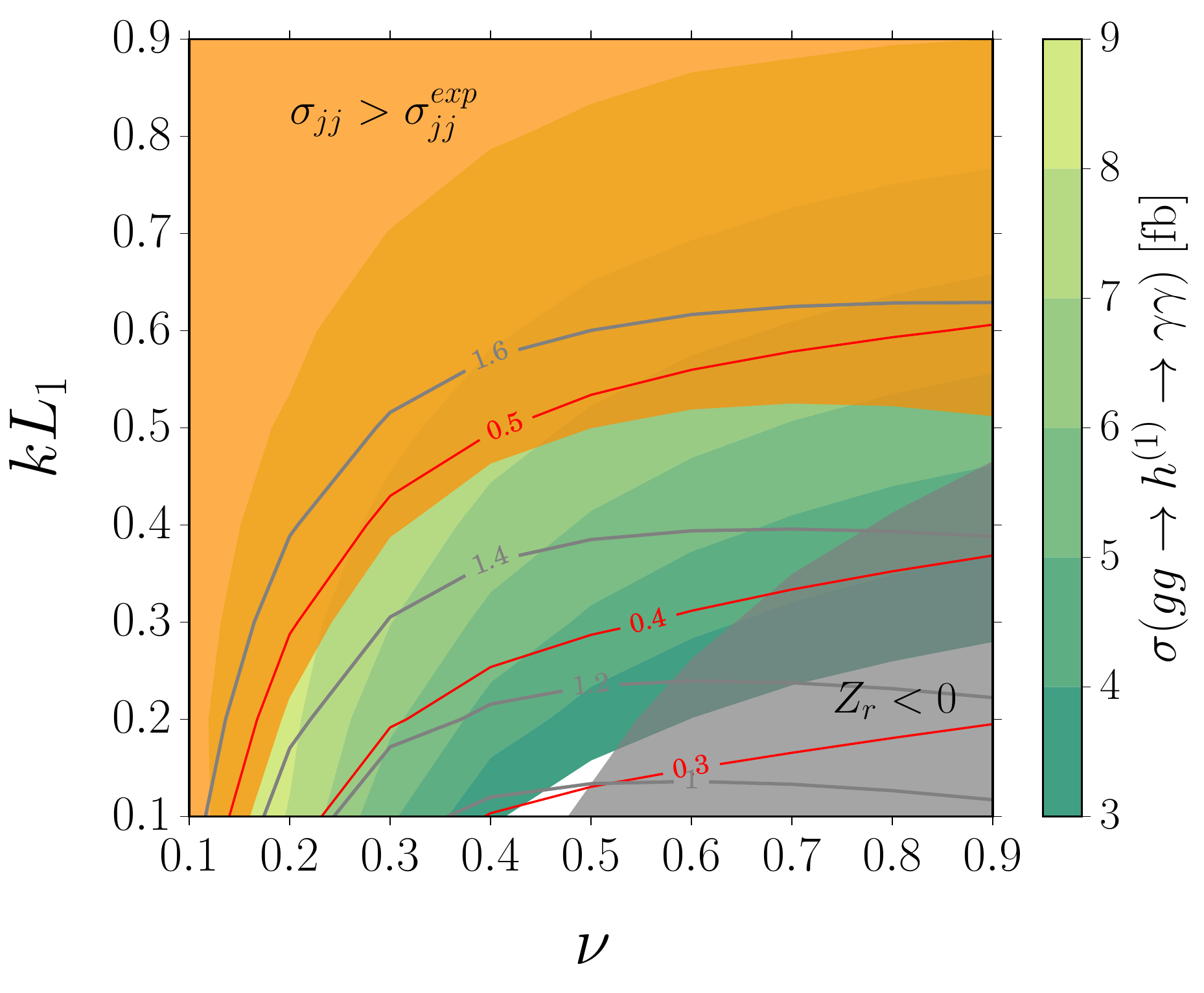}
	\caption{Contour values of  $\sigma(gg\to h^{(1)}\to \gamma\gamma)$  in the $\nu-k L_1$ plane together with the exclusion bounds arising from di-jet searches (orange) and from having a radion ghost (grey). We also show in red contour lines for $M_5 L_1\in\{0.3,\,0.4,\, 0.5\}$. We have assumed $m_{\rm KK}=0.9\, \TeV$, $A(y_1)=37.5$ and set $\tilde{\kappa}=2.2$.  $t\bar{t}$ searches are not competitive enough to constraint this region of the parameter space. For completeness, we also display (in grey) several contour lines for the ratio $M_5/\bar{M}_{\rm Pl}$. }
		\label{fig:xsec0p9TeV}
\end{figure}

\begin{figure*}[!t]
	\centering
	\includegraphics[width=\textwidth]{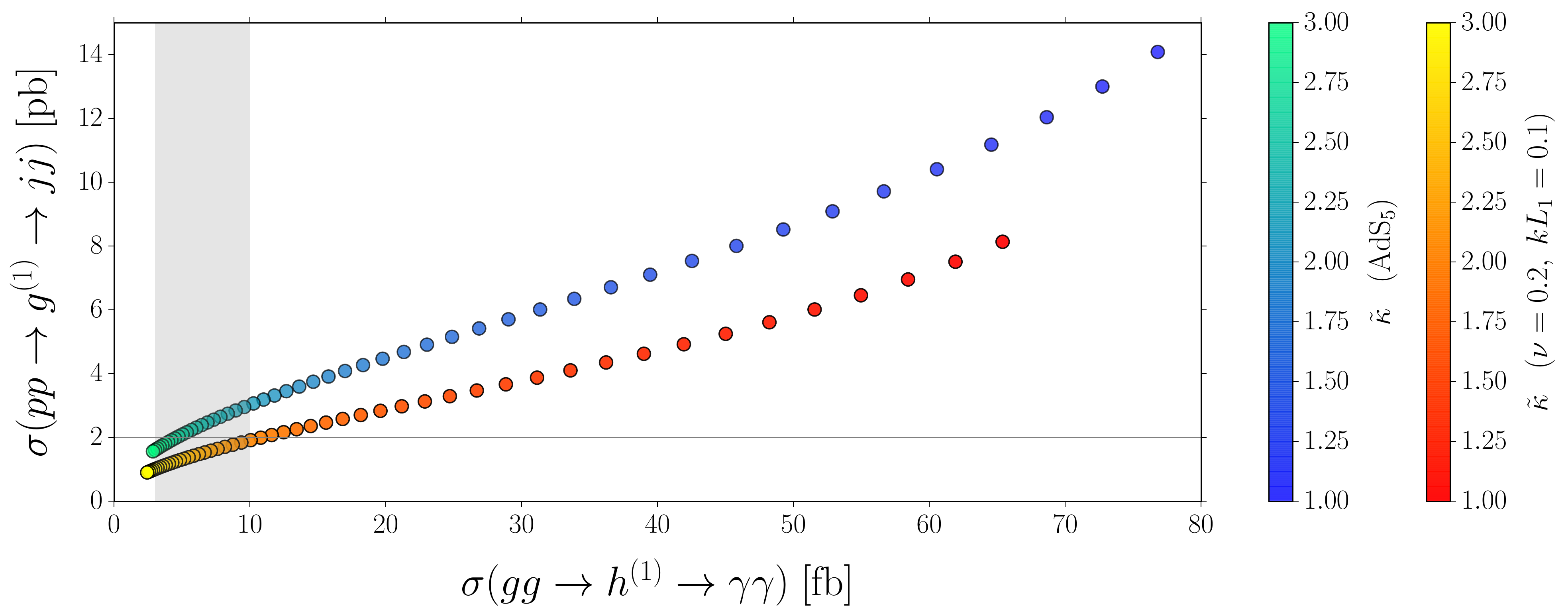}
	\caption{Di-jet cross section $\sigma(pp\to g^{(1)}\to jj)$ and the KK-graviton di-photon cross section $\sigma(gg\to h^{(1)}\to \gamma\gamma)$ as a function of $\tilde{\kappa}\in[1,3]$ for $\nu=0.2,\, kL_1=0.1$ and the AdS$_5$ case. In both cases we have assumed $A(y_1)=37.5$ and $m_{\rm KK}=0.9\,\TeV$. The horizontal grey line correspond to the upper bound on di-jet production, whereas the vertical grey band corresponds to a di-photon cross section of $[3,\, 10]$\, fb.}
		\label{fig:kappa}
\end{figure*}

In the setup at hand, the KK-graviton couples mostly to gluons and electroweak gauge bosons,  leading therefore to di-photon production  via gluon fusion, $gg\to h^{(1)}\to \gamma \gamma$, which is favored compared to other production mechanisms when one takes into account the 8 TeV data \cite{Franceschini:2015kwy}.  According to the current experimental data, a total cross section of $\sigma(gg\to h^{(1)}\to \gamma\gamma)\sim 5~\fb$ is required in order to accommodate the observed anomaly. On the other hand, the strongest constraint due to the presence of the vector KK spectrum in these setups is di-jet production \cite{CMS-PAS-EXO-14-005, Aad:2014aqa, Khachatryan:2015sja} via  the $s$-channel exchange of the KK gluon $pp\to g^{(1)}\to jj$. We assumme a QCD K-factor $\kappa_{qqg^{(1)}}=1.3$ \cite{Gao:2010bb}. The presence of electroweak vector resonances do not lead to significant collider constraints since they decay almost $100\%$ of the time  to the dark scalars, for they have volume enhanced couplings since they all come from the strongly interacting sector \cite{Carmona:2015haa}. This could be also the case for the KK graviton but, since we are forced to consider $m_{\rm KK}> m_{h}^{(1)}$, we will assume that the pair production of dark scalars is not kinematically open for the spin-2 resonance, i.e. $m_{h}^{(1)}/2<m_{\pi}< m_{\rm KK}/2$. Since the masses of the pNGB are linked to the KK scale $m_{\rm KK}$, this will be always true for moderately large values of the latter. Otherwise, additional sources of breaking of the Goldstone symmetry would be required. We have also considered the bounds arising from $t\bar{t}$ production \cite{Chatrchyan:2013lca}. Note that due to the IR localization of the KK graviton, its di-lepton cross section will be much smaller than $\sigma(gg\to h^{(1)}\to \gamma\gamma)\sim 5\, \fb$ and therefore well beyond current experimental sensitivity. In Figure\,\ref{fig:xsec0p9TeV}, we display contour values in the $\nu-kL_1$ plane for the di-photon cross section $\sigma(gg\to h^{(1)}\to \gamma\gamma)$ together with the excluded regions arising from di-jet searches (orange) and the presence of a radion ghost (grey), for $m_{\rm KK}=0.9\,$TeV, $A(y_1)=37.5$ and $\tilde{\kappa}=2.2$. We also show in red contour lines for $M_5 L_1\in\{0.3,\, 0.4,\, 0.5\}$, to better assess the preferred value of $M_5 L_1$ giving the desired di-photon cross section. Finally, we also display for completeness some contour lines (in grey) for the ratio $M_5/\bar{M}_{\rm Pl}$.  We have explicitly checked that $t\bar{t}$ searches are not competitive enough to constraint this region of the parameter space. All these processes have been computed at the parton level using \texttt{MadGraph~v5} \cite{Alwall:2014hca} after implementing the model via \texttt{Feynrules~v2} \cite{Alloul:2013bka}. 

\begin{figure}[!h]
	\centering
	\includegraphics[width=\columnwidth]{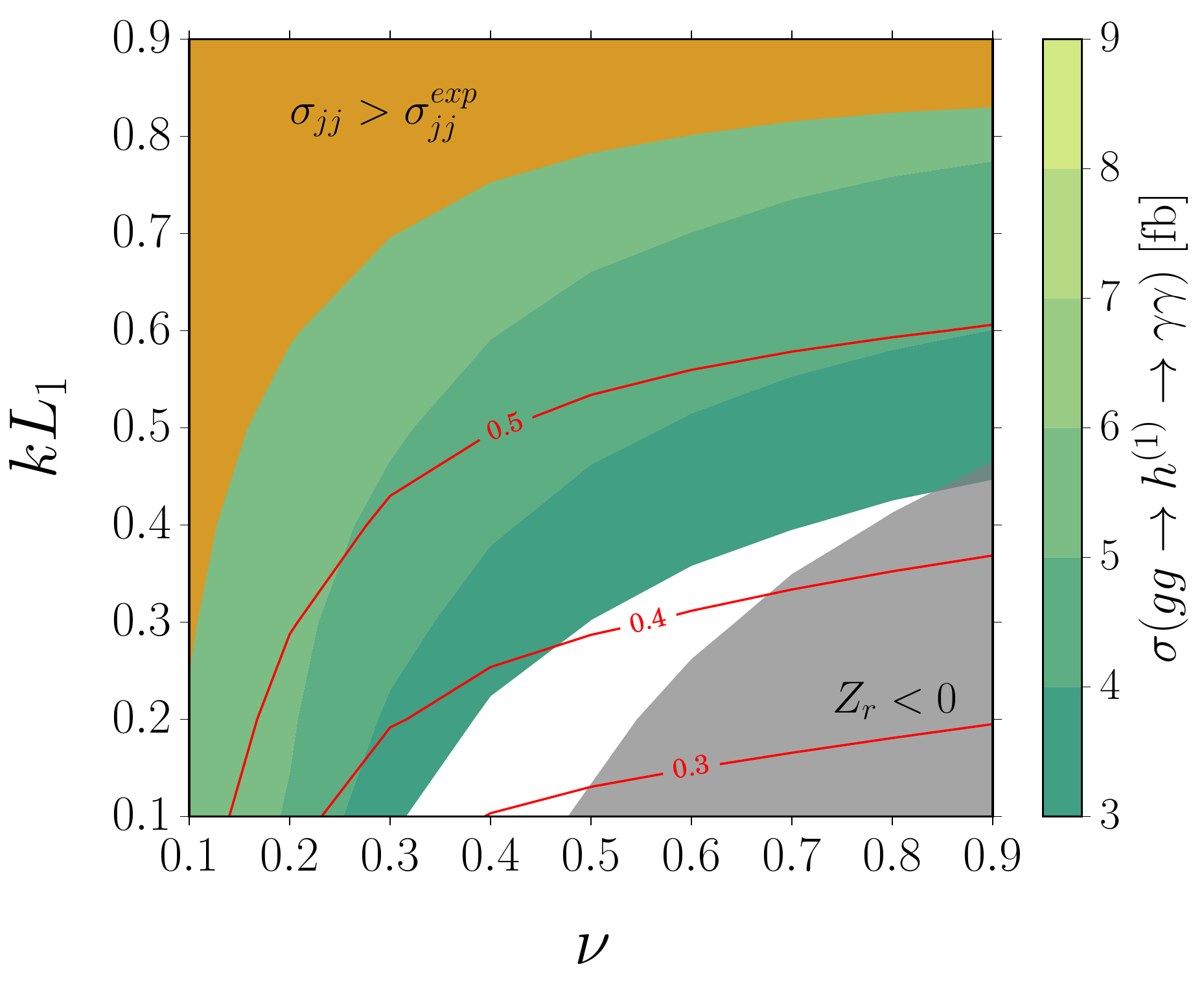}
	\caption{Contour values of  $\sigma(gg\to h^{(1)}\to \gamma\gamma)$  in the $\nu-k L_1$ plane together with the exclusion bounds arising from di-jet searches (orange) and from having a radion ghost (grey). We also show in red contour lines for $M_5 L_1\in\{0.3,\,0.4,\, 0.5\}$. We have assumed $m_{\rm KK}=0.9\, \TeV$, $A(y_1)=37.5$ and set $\tilde{\kappa}=2.5$.  }
		\label{fig:xsec0p9TeV_2}
\end{figure}

In Figure\,\ref{fig:xsec0p9TeV}, we have chosen the minimal value of $\tilde{\kappa}$ that maximizes the allowed region in the $\nu-kL_1$ plane. Since the couplings $ggh^{(1)}$ and $\bar{q}qg^{(1)}$ scale with $1/\tilde{\kappa}^2$ and $1/\tilde{\kappa}$, respectively, the corresponding production cross sections will be $\sigma(pp\to g^{(1)})\propto 1/\tilde{\kappa}^{2}$ and $\sigma(gg\to h^{(1)})\propto 1/\tilde{\kappa}^4$. On the other hand, the couplings of the KK-graviton to the electroweak gauge bosons do not depend on $\tilde{\kappa}$, making $\mathrm{BR}(h^{(1)}\to \gamma\gamma)\propto 12\tilde{\kappa}^4/(8+4\tilde{\kappa^4})$ with good approximation, whereas $\mathrm{BR}(g^{(1)}\to jj)$ will remain $\approx 5/6$. Therefore, increasing the value of $\tilde{\kappa}$ reduces the di-photon cross section $1/\tilde{\kappa}^2$ faster than the di-jet one, modulo a factor 3 that can be gained via the enhanced $\mathrm{BR}(h^{(1)}\to \gamma\gamma)$ for large $\tilde{\kappa}$. To study this effect in more detail we show in Figure\,\ref{fig:kappa} the aforementioned cross sections as a function of $\tilde{\kappa}$ for the AdS$_5$ case and $\nu=0.2,\, kL_1=0.1$, assuming $m_{\rm KK}=0.9\,\TeV$ and $A(y_1)=37.5$. For this particular point of the $\nu-kL_1$ plane and $\tilde{\kappa}=1$, the ratio in question is $\sim 1.5$ times bigger than the one obtained in the AdS$_5$ case.  This is due to the fact that deformations of conformality have a bigger impact on the $\bar{q}q g^{(1)}$ coupling than in the $ggh^{(1)}$ one, reducing the former slightly  more than the latter. Since greater values of $\tilde{\kappa}$ will decrease the di-photon cross section faster than the di-jet one, this effect will be very valuable in order to fulfil current experimental bounds and at the same time reproduce the di-photon excess, as can be seen from Figure\,\ref{fig:kappa}.\,\footnote{One should note however that, for constant values of $M_5L_1$, larger values of $A(y_1)$ will be allowed for points with a smaller deformation of conformality. Therefore, for fixed values of $M_5L_1$ and $m_{\rm KK}$, the effect just mentioned will be  compensated to some extent by the increase in $A(y_1)$, which tends to enhance the $ggh^{(1)}$ coupling with a much smaller effect on $\bar{q}qg^{(1)}$.}  At any rate, from this Figure one can readily conclude that the AdS$_5$ case is also  allowed  for $m_{\rm KK}=0.9\,\TeV$ and $A(y_1)=37.5$. In order to assess further the impact of $\tilde{\kappa}$ on the parameter space, we show in Figure\,\ref{fig:xsec0p9TeV_2} the equivalent  of Figure\,\ref{fig:xsec0p9TeV} for a larger value of $\tilde{\kappa}=2.5$. We can see that regions with a smaller deformation of conformality are now preferred, even though smaller values for the di-photon cross section are obtained.  

\begin{figure}[!ht]
	\centering
	\includegraphics[width=\columnwidth]{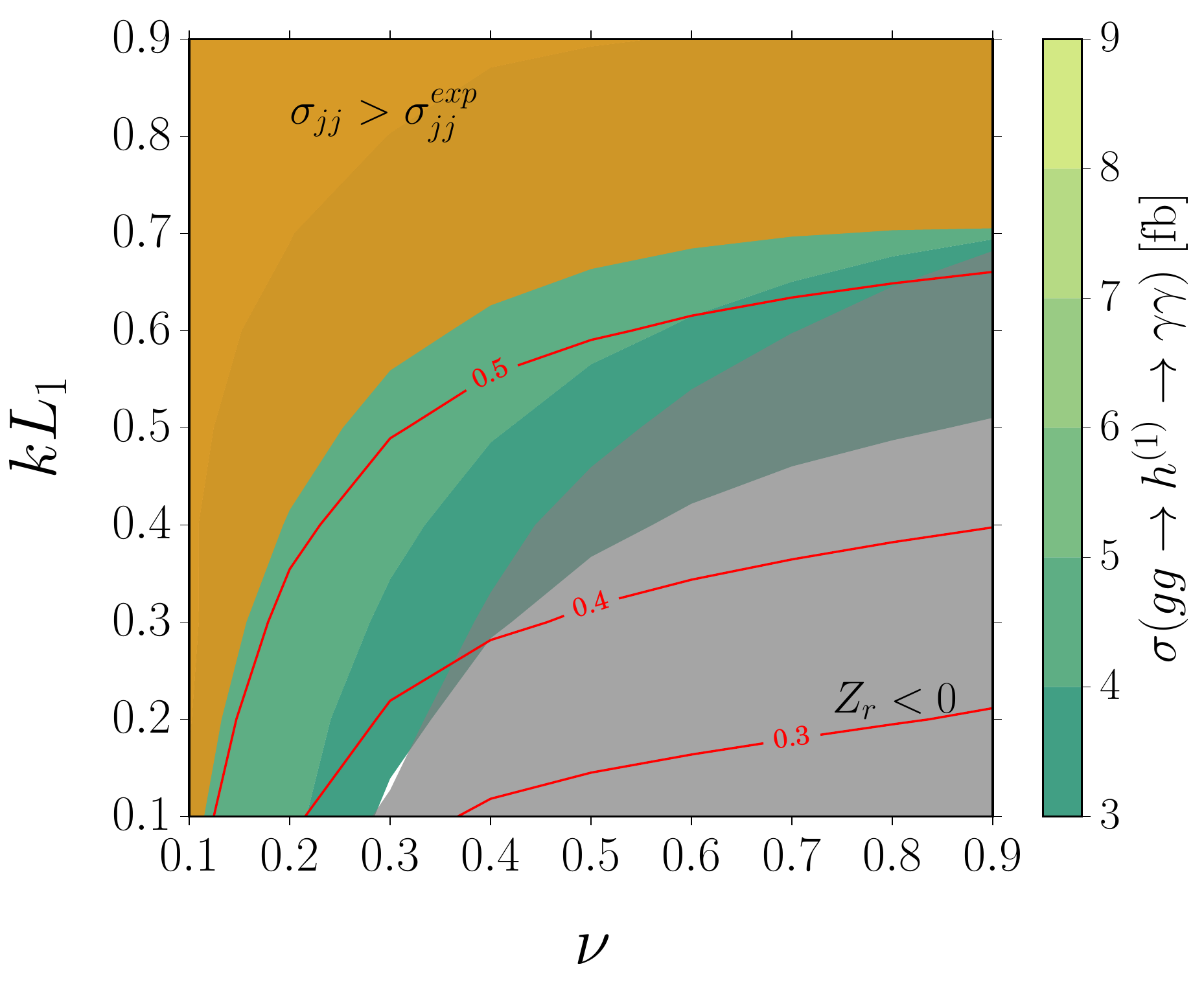}
	\caption{Contour values of  $\sigma(gg\to h^{(1)}\to \gamma\gamma)$  in the $\nu-k L_1$ plane together with the exclusion bounds arising from di-jet searches (orange) and from having a radion ghost (grey). We also show in red contour lines for $M_5 L_1\in\{0.3,\,0.4,\, 0.5\}$. We have assumed $m_{\rm KK}=0.95\, \TeV$, $A(y_1)=37.5$ and set $\tilde{\kappa}=2.5$. }
		\label{fig:xsec0p95TeV}
\end{figure}
\begin{figure}[!ht]
	\centering
	\includegraphics[width=\columnwidth]{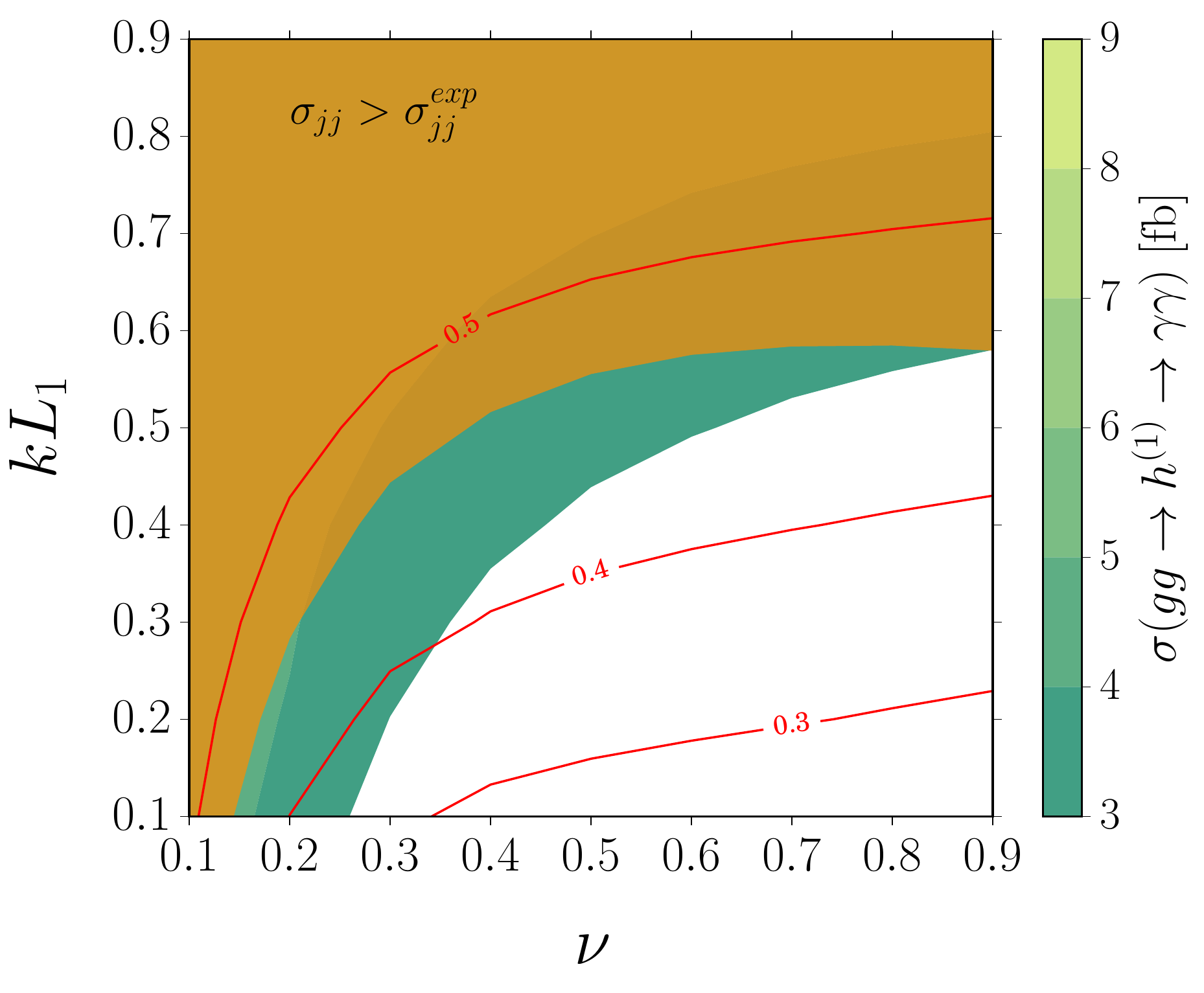}
	\caption{Contour values of  $\sigma(gg\to h^{(1)}\to \gamma\gamma)$  in the $\nu-k L_1$ plane together with the exclusion bounds arising from di-jet searches (orange). We also show in red contour lines for $M_5 L_1\in\{0.3,\,0.4,\, 0.5\}$. We have assumed $m_{\rm KK}=1.0\, \TeV$, $A(y_1)=37.5$ and set $\tilde{\kappa}=2.5$. Note that the whole region is in principle excluded by the presence of a radion ghost.}
		\label{fig:xsec1TeV}
\end{figure}

Increasing $m_{\rm KK}$ to $0.95\,\TeV$ leads to a slightly smaller di-photon cross section, which moderately reduces the allowed region in the $\nu-kL_1$ plane, as can be seen from Figure\,\ref{fig:xsec0p95TeV}, where we show again the contour values of $\sigma(gg\to h^{(1)}\to \gamma \gamma)$ together with the di-jet and $Z_r<0$ excluded regions, for $m_{\rm KK}=0.95\,\TeV$, $A(y_1)=37.5$ and $\tilde{\kappa}=2.5$.  Again, we display in red contour lines for $M_5 L_1\in \{0.3,\, 0.4,\, 0.5\}$. Note that there is a larger exclusion region coming from the presence of a radion ghost, which nevertheless does not overlap significantly with the area leading to the correct di-photon cross section. Further increasing $m_{\rm KK}$ to $1\,\TeV$  leads to a notable  decrease of the allowed parameter space for $A(y_1)=37.5$ and $\tilde{\kappa}=2.5$, as it is shown in Figure\,\ref{fig:xsec1TeV}. Note however that this value of $m_{\rm KK}$ is anyhow excluded by the presence of a radion ghost.

\section{Discussion and Conclusions}
\label{sec:conc}
We have shown that the $750\,\GeV$ resonance, if experimentally confirmed, can be the KK graviton of an approximately conformal dark sector, which accounts for the bulk of the observed DM relic abundance. In these setups, the KK graviton couples universally to all SM gauge bosons (modulo possible gauge kinetic terms) and have negligible couplings to the rest of the SM.  We have explicitly shown that the masses of the vector resonances can not be taken arbitrarily large if one wants to have perturbativity in the 5D gravitational theory and at the same time explain the di-photon anomaly. Moreover, if avoiding the presence of a radion ghost for large KK masses proves to be unachievable or it comes at the price of a large phenomenological impact, the resultant upper bound $m_{\rm KK}\lesssim 1\,\TeV$ would be a strong case for these scenarios since, contrary to other setups explored recently in the literature \cite{Han:2015cty, Giddings:2016sfr, Sanz:2016auj, Falkowski:2016glr, Liu:2016mpd, Hewett:2016omf}, they can feature light enough vector resonances without any theoretical or experimental problem. Indeed, a robust prediction in these scenarios is the presence of a $\mathcal{O}(1)\,\TeV$ color octet  resonance with universal coupling to fermions, which are probed essentially by di-jets searches. Since the strongly interacting dark sector plays no role in EWSB, light  electroweak vector resonances can be present without contradicting EWPT and collider searches, provided they decay dominantly to the dark scalars, which is a natural expectation in these models.  Our setup also provides a very concrete prediction for the dark scalar masses, since the KK-graviton should not be allowed to pair produce them,  $m_{h}^{(1)}/2=375\,\GeV \lesssim m_{\pi} \lesssim  500\,\GeV$. In summary, we have  presented the first phenomenological study of setups providing a $750\,\GeV$ spin-2 excitation where the effect of the vector resonances is not decoupled, motivating it by the consistency of the five dimensional theory and exploring in detail the interplay between all experimental and theoretical constraints.

\acknowledgments
I would like to thank Florian Goertz for fruitful discussions that initialized the project. I would also like to thank Kaustubh Agashe, Jose Santiago and Mikael Chala for useful comments and discussion.   This research has been supported by a Marie Sk\l{}odowska-Curie Individual Fellowship of the European Community's Horizon 2020 Framework Programme for Research and Innovation under contract number 659239 (NP4theLHC14).

\newpage
\bibliography{gravitons}
\bibliographystyle{apsrev4-1}

\end{document}